\documentclass[12pt,preprint]{aastex}
\usepackage{booktabs}
\newcommand{\Msun}{~M_\odot}

\newcommand{\lsim}{\raise0.3ex\hbox{$<$}\kern-0.75em{\lower0.65ex\hbox{$\sim$}}}
\newcommand{\gsim}{\raise0.3ex\hbox{$>$}\kern-0.75em{\lower0.65ex\hbox{$\sim$}}}
\newcommand{\propsim}{\raise0.3ex\hbox{$\propto$}\kern-0.75em{\lower0.65ex\hbox{$\sim$}}}

\newcommand{\Mdot}{\dot M}

\begin{document}
    
\title{The radio spectra of SN\,2020oi: Effects of radiative cooling on the deduced source properties}

\author{C.-I. Bj\"ornsson\altaffilmark{1}}
\altaffiltext{1}{Department of Astronomy, AlbaNova University Center, Stockholm University, SE--106~91 Stockholm, Sweden.}
\email{bjornsson@astro.su.se}

\begin{abstract}
Observations of radiative cooling in a synchrotron source offer a possibility to further constrain its properties. Inverse Compton cooling is indicated in the radio spectra during the early phases of SN\,2020oi. It is shown that contrary to previous claims, observations are consistent with equipartition between relativistic electrons and magnetic field as well as a constant mass-loss rate of the progenitor star prior to the supernova explosion. The reason for this difference is the need to include cooling directly in the fitting procedure rather than estimating its effects afterward. It is emphasized that the inferred properties of the supernova ejecta are sensitive to the time evolution of the synchrotron self-absorption frequency; hence, great care should be taken when modeling spectra for which cooling and/or inhomogeneities are indicated. Furthermore, it is noted that the energies of the relativistic electrons in the radio emission regions in supernovae are likely too low for first-order Fermi acceleration to be effective.
\end{abstract}


\section{Introduction}
Most of the energy emitted by supernovae falls in the optical/infrared spectral regime and is of thermal origin. Through spectral diagnostic, density and temperature of the emitting gas can be determined and then used to infer properties of both the supernova ejecta and the circumstellar medium. The radiated energy coming out as radio emission is only a small fraction of the total. However, due to its non-thermal nature, it can probe aspects of the supernova explosion as well as the progenitor star not easily achieved by optical/infrared observations.

The radio emission is thought to originate in between the forward and reverse shocks, which result from the interaction between the supernova ejecta and the surrounding wind from the progenitor star. Standard synchrotron theory gives estimates of energy densities in magnetic field and relativistic particles, which set a lower limit to the thermal energy density. The energy densities, derived from radio observations compiled by \cite{bie21}, have been shown to significantly constrain various mass-loss scenarios for the progenitor stars \citep{mor21,m/y22}. Another example is the use of the outer radius of the radio-emitting region, which is usually taken to correspond to the forward shock. Since the source radius can be deduced from the synchrotron self-absorption frequency, in well-observed supernovae the evolution of the forward shock can be followed. The dynamics of the forward shock is determined by the density structure of the ejecta; for example, its radial gradient is directly related to the deceleration of the forward shock. Standard synchrotron modeling of the initial phases of SN\,1993J \citep{f/b98} is not consistent with explosion models developed to account for the optical/infrared emission \citep{bjo15}. Even the most favorable ones fall short by more than an order of magnitude to provide the kinetic energy indicated by radio observations at high ejecta velocities. Also, a consistent description of the radio emission can be used to distinguish between viable models for the X-ray emission.

Furthermore, radio observations have implications that go beyond the understanding of the supernovae themselves.  Acceleration of particles and amplification of magnetic fields are both issues of relevance for a large number of astrophysical phenomena. With the advent of high-quality radio spectra taken over extended periods of time, supernova observations give probably the best opportunities to study one of the sites where these mechanisms are at work; in particular, the short evolutionary timescale for supernovae allows following the effects of a changing environment. Another advantage over, for example, compact radio sources/blazars \citep{bla19} is that the general setting is better constrained in supernovae than in these extragalactic objects. Together with the recent development of a kinematical description of shock formation and the associated acceleration of particles and magnetic field amplification \citep{c/s14a,c/s14b}, radio supernovae hold the promise for a better understanding of the still rather unknown underlying physics.

Therefore, the potential of radio supernovae to contribute to the understanding of a range of phenomena is large. However, in order to be realized, the synchrotron modeling itself needs to be secure. The deduction of source parameters from a standard analysis of a self-absorbed synchrotron spectrum relies on various assumptions. Hence, sources deserve special attention, for which additional constraints are available. Although not common, there are a number of sources where either radiative cooling or inverse Compton scattered X-ray radiation is indicated. Even so, the analysis is not straightforward, since the value of the cooling frequency is often hard to constrain and the use of the observed X-ray radiation is limited by the effects of inhomogeneities \citep{bjo13}. 

Radio and X-ray observations of SN\,2011dh have been used to argue that conditions in the synchrotron emitting region is far from equipartition between relativistic electrons and magnetic fields \citep{sod12,kra12,hor13}. A similar conclusion has been reached for SN\,2012aw \citep{yad14} and SN\,2013df \citep{kam16} based on the indicated presence of inverse Compton cooling. In SN\,2002ap both Compton scattered X-ray and Compton cooling are suggested by observations \citep{ber02,sut03}. Here, instead, \cite{b/f04} find that observations are consistent with equipartition between relativistic electrons and magnetic field. This is surprising, since the shock environment is not expected to differ greatly between these stripped-envelope supernovae. Hence, if true, this suggests that the mechanisms responsible for acceleration of particles and amplification of the magnetic field are quite sensitive to local conditions. 

This is one example of a central question that can be addressed with supernova observations; therefore, it is important to establish whether the outcome of an analysis is real or likely due to the modeling procedure. The effects of inverse Compton scattering are not restricted to stripped-envelope supernovae but are observed also in other supernovae, for example, in a few Type IIPs, SN\,2004dj \citep{cha12,nay18} and  SN\,2016X \citep{r-c22}. 

The main focus of the present paper is to discuss the treatment of radiative cooling in a synchrotron source. A reanalysis is then made of the detailed radio observations of Type Ic SN\,2020oi done by \cite{hor20}. This paper is structured as follows: the problem of deducing the cooling frequency from observations is illustrated in Section\,2 together with a few other issues relevant to the analysis. In Section\,3, an alternative way of analyzing the observations of SN\,2020oi is introduced. It is argued that this gives more reliable constraints on the source parameters. As discussed in Section\,4, and contrary to the conclusions in \cite{hor20}, the observations are consistent with equipartition as well as a constant mass-loss rate of the progenitor star. It is also emphasized that the structure of the ejecta is determined mainly by the decline rate of the self-absorption frequency. Hence, systematic effects in the fitting procedure can seriously affect the deduced properties of the ejecta. The conclusions of the paper are presented in Section\,5. Numerical results are mostly given using cgs units. When this is the case, the units are not written out explicitly.

\section{Deducing Source Properties from the Observed Synchrotron Emission}\label{sect2}
In order to deduce source properties from observations, a model is needed. The simplest assumption is that of a homogeneous, spherically symmetric source. However, even such a simple model cannot always be uniquely constrained by observations. In most cases, the observables are limited to the self-absorption frequency ($\nu_{\rm abs}$) and the corresponding spectral flux ($F_{\nu_{\rm abs}}$) in addition to the optically thin spectral index ($\alpha$). The radiating electrons are usually taken to have a distribution of Lorentz factors $(\gamma$) given by $n(\gamma) = K_{\rm o} \gamma^{-p}$ for $\gamma > \gamma_{\rm min}$ and $p\,>\,2$, so that $\alpha =(p-1)/2$. The magnetic field strength ($B$) can then be expressed as
\begin{equation}
	B = 1.0\frac{\nu_{{\rm abs},10}}{\left(y^2 F_{\nu_{\rm abs},27}\right)^{2/19}},
	\label{eq5}
\end{equation}
and the source radius is
\begin{equation}
	R = 5.5\times 10^{15}\frac{F_{\nu_{\rm abs},27}^{9/19}}{y^{1/19}\nu_{\rm abs,10}},
	\label{eq6}
\end{equation}
where $p\,=\,3$ has been assumed. The expressions for arbitrary $p$ can be found in \cite{bjo21}. Here, $\nu_{{\rm abs},10} \equiv \nu_{\rm abs}/10^{10}$ and $F_{\nu_{\rm abs},27} \equiv F_{\nu_{\rm abs}}/10^{27}$. Furthermore, $\nu_{\rm abs}$ is defined as the frequency where the spectral flux peaks.

The expression for $y$ is given by
\begin{equation}
	y = \gamma_{\rm min}\frac{U_{\rm rel}}{U_{\rm B}}\frac{R_{||}}{R},
	\label{eq7}
\end{equation}
where $R_{\rm ||}$ is the average line-of-sight extension of the source, $U_{\rm rel}$ and $U_{\rm B}$ are the energy densities of the relativistic electrons and the magnetic field, respectively. Hence, it is seen that the value of $y$ involves a threefold degeneracy, two of which are related to poorly understood physics, namely, the injection of particles into the acceleration process ($\gamma_{\rm min}$) and the partition of energy between relativistic electrons and magnetic fields ($U_{\rm rel}/U_{\rm B}$). The emission is normally assumed to come from a thin shell behind an expanding forward shock. For a spherically symmetric source, the value of $R_{||}$ is twice the thickness of the emitting shell. For a strong shock, this implies $R_{||}/R = 1/2$.

In cases when only $F_{\nu_{\rm abs}}$ and $\nu_{\rm abs}$ are available (in addition to $p$), an assumption on the value of $y$ has to be made in order to estimate the radius of the source and its magnetic field. A third observable is needed to constrain the value $y$. When X-ray emission is detected and it can be argued that it is due to inverse Compton scattering by the same electrons, which give rise to the radio emission, a value for $y$ can be obtained. Let $L_{\rm x}$ and $L_{\rm radio}$ be the X-ray and radio luminosities, respectively, where the luminosity is defined as $L = \nu F_{\nu}$. For a homogeneous source $L_{\rm x}/L_{\rm radio} = U_{\rm ph}/U_{\rm B} = 2 L_{\rm bol}/cR^2B^2$, where $L_{\rm bol}$ is the luminosity of the seed photons for the inverse Compton scattering. Equations (\ref{eq5}) and (\ref{eq6}) directly give an expression for $RB$ and one finds
\begin{equation}
	y = 0.23\,F_{\nu_{\rm abs, 27}}^{7/5}\left(\frac{L_{\rm x}}{L_{\rm radio} L_{\rm bol, 42}}\right)^{19/10},
	\label{eq8}
\end{equation}
where $L_{\rm bol, 42} \equiv L_{\rm bol}/10^{42}$. 

\subsection{Determining the Cooling Frequency in an Optically Thin Synchrotron Source} \label{sect2a}
Another effect that can be used to constrain the source parameters is radiative cooling. In an expanding, spherically symmetric source the column density of relativistic electrons injected behind the forward shock is \citep{f/b98}
\begin{equation}
	N(\gamma) = \frac{K_{\rm o}vt}{({\rm p} -1)\gamma^{\rm p}}\frac{1}{(1+t/t_{\rm cool})}.
	\label{eq1}
\end{equation}
Here, $v$ is the velocity of the plasma behind the shock and $t$ is the time since the beginning of the expansion. The cooling time for either inverse Compton or synchrotron radiation can be written $t_{\rm cool} = 1/a\gamma$ with $a = 4\sigma_{\rm T}U/(3mc)$, where $U$ is the energy density of either photons ($U_{\rm ph}$) or magnetic fields ($U_{\rm B}$) and $\sigma_{\rm T}$ is the Thomson cross section. 

It can be seen from Equation (\ref{eq1}) that the transition from the non-cooling part (i.e., $\gamma \ll 1/at$) to the cooling part (i.e., $\gamma \gg 1/at$) is quite extensive. This can be illustrated by calculating the variation of the column density with $\gamma$, i.e., $\hat{{\rm p}} = - {\rm d}\ln N(\gamma)/{\rm d}\ln \gamma$. This gives
\begin{equation}
	\hat{{\rm p}} = {\rm p} + \frac{a \gamma t}{1+ a\gamma t}.
	\label{eq2}
\end{equation}
The cooling frequency is $\nu_{\rm  cool} \propto \gamma_{\rm cool}^2$, where $\gamma_{\rm cool} = 1/at$, so that $\hat{{\rm p}} = {\rm p}+ 1/2$ at $\nu = \nu_{\rm cool}$. To some extent, cooling affects the spectrum over the whole frequency range; for example, since $a\gamma t = (\nu/\nu_{\rm cool})^{1/2}$, changing $\hat{p}$ by half its total amount (i.e. $\Delta \hat{p} = 0.5$) requires a frequency range of $\sim10^2$ even around $\nu = \nu_{\rm cool}$.

When cooling is not important, the spectral flux is $F_{\nu} \propto \nu^{-(p-1)/2}$. Since the range of $\gamma$-values contributing to the flux at a given frequency is quite small, a good approximation for the local spectral flux is obtained by substituting $\hat{p}$ for $p$, i.e., $F_{\nu} \propto \nu^{-(\hat{p}-1)/2}$, which gives a local spectral index $\alpha = (\hat{p} - 1)/2$. In principle, the cooling frequency could then be determined directly from the observed spectrum. However, in practice, this meets with some limitations: (1) Due to the slow variation of $\alpha$ with frequency, a wide frequency range is needed in order to determine $p$. (2) A high-quality spectrum is required to determine the local spectral index with accuracy enough to meaningfully constrain the cooling frequency. 

The importance of adiabatic cooling for smearing out the transition between the non-cooling and cooling parts of the spectrum can  be appreciated by comparing it to a situation where only radiative cooling is included. Consider electrons injected with a Lorentz factor $\gamma_{\rm o}$. After a time $\hat{t}$, their Lorentz factor is $\gamma = \gamma_{\rm o}/(1+a\gamma_{\rm o}\hat{t})$ (or $\gamma_{\rm o} = \gamma/(1-a\gamma \hat{t})$). Hence, ${\rm d} \gamma_{\rm o}/{\rm d} \gamma = 1/(1-a\gamma \hat{t})^2$. With the same distribution of injected energies as above, the density of electrons with Lorentz factor $\gamma$ is
\begin{equation}
	n(\gamma) =\frac{K_{\rm o}(1-a\gamma \hat{t})^{p-2}}{\gamma^p}.
	\label{eq3}
\end{equation}
At a time $t$, the column density is obtained by integrating $v n(\gamma)$ from $\hat{t} = 0$ to $\hat{t} = {\rm min}[t, 1/a\gamma]$. The result is
\begin{eqnarray}
	N(\gamma) &=& \frac{K_{\rm o} v}{(p-1)a \gamma^{p+1}}\left[1 - (1-a\gamma t)^{p-1}\right] , \hspace{1cm}t< 1/a\gamma \nonumber \\
	&=& \frac{K_{\rm o} v}{(p-1)a \gamma^{p+1}} , \hspace{4.5cm}t> 1/a\gamma	
	\label{eq4}
\end{eqnarray}

It can be seen that for $\gamma > 1/a t$ the result corresponds to the limit $t \to \infty$ when adiabatic cooling is included. The width of the transition region for $\gamma < 1/a t$ increases with the value of $p$; it may be noted that for $p=2$, it is zero, i.e., there is an abrupt change from $ \hat{p} = p$ to $\hat{p} = p +1$ at $\gamma = 1/at$. In order to compare with the above result for the adiabatic cooling case, where the frequency variation of $\alpha$ is independent of $p$, let $p=3$. It can be deduced from Equation (\ref{eq4}) that $\Delta \hat{p} = 0.5$ just below $\nu_{\rm cool}$, then occurs over a frequency range of $\sim 2$. Hence, the transition region is much narrower than when adiabatic cooling is included. For such cases, the cooling frequency would be considerably easier to determine from observations.

\subsection{The Effects of Cooling on the Synchrotron Self-absorption Frequency}\label{sect2b}
Radiative cooling affects the value for $y$, since $R_{||}/R$ becomes smaller by a factor $1+t/t_{\rm cool}$ (see Equation (\ref{eq1})). When cooling is dominated by synchrotron emission, the value of $B$ is obtained directly from $t/t_{\rm cool}$, and hence, $y$ can be deduced from Equation (\ref{eq5}). On the other hand, inverse Compton cooling implies
\begin{equation}
	\frac{t}{t_{\rm cool}} = 3.6\times10^{-2}\frac{\nu_{10}^{1/2}L_{42}t_{10}}{B^{1/2}R_{16}^2},
	\label{eq9}
\end{equation}
where $t_{10} \equiv t$/10\,days, $\nu_{10} \equiv \nu/10^{10}$ and $R_{16} \equiv R/10^{16}$. Furthermore, $\nu = 1.6\,\gamma^2 \nu_{\rm B}$ (see below) has been used, where $\nu_{\rm B}$ is the cyclotron frequency. The expressions for $B$ and $R$ in Equations (\ref{eq5}) and (\ref{eq6}) then yield
\begin{equation}
	y^{4/19}L_{42} = 8.4\frac{t}{t_{\rm cool}}\frac{1}{\nu_{10}^{1/2}t_{10}}\frac{F_{\nu_{\rm abs, 27}}^{17/19}}
	{\nu_{\rm abs, 10}^{3/2}}.
	\label{eq10}
\end{equation}

When both radiative cooling and the Compton scattered X-ray emission are observed, it is possible to test for the validity of the basic underlying assumption of a homogeneous source structure. If the observed X-ray emission is larger than predicted from the analysis of the radio observations, this indicates that only a fraction of the relativistic electrons are located within the synchrotron emitting volume, i.e., the relativistic electrons occupy a larger volume than does the magnetic field. Likewise, when the Compton scattered X-ray emission is observed but no radiative cooling, the value of $y$ deduced from Equation (\ref{eq8}) should be treated with some care. If the source is inhomogeneous, the value obtained will be artificially enhanced. This shows that a reliable value of $U_{\rm rel}/U_{\rm B}$ is hard to obtain. Furthermore, observations give estimates only of the energy density of the electrons radiating in the radio regime (i.e., $U_{\rm rel}(\gamma)$). Since $\gamma_{\rm min} U_{\rm rel} \approx \gamma U_{\rm rel}(\gamma)$, it is not possible to separate the values of $\gamma_{\rm min}$ and $U_{\rm rel}$. 

\subsection{Connection between the Observed Frequency and Electron Energy}\label{sect2c}
The spectral distribution of the synchrotron radiation emitted by a single electron is given by $F(z)$ \citep{r/l04,tuc75}, which peaks at $z=0.29$. Here, $z = \nu/\nu_{\rm c}$, where $\nu_{\rm c} = (3/2)\sin \theta \gamma^2 \nu_{\rm B}$ and $\theta$ is the pitch angle. For an isotropic electron distribution, this gives for the peak frequency $\nu = 0.34\,\gamma^2 \nu_{\rm B}$. A power-law distribution of electron energies results in a spectral emissivity $j_{\nu} \propto \int z^{(p-3)/2} F(z) {\rm d} z$. Hence, the average value of $z$ depends on $p$, for example, $<z(p=3)>\,= 1.32$ and $<z(p=2)>\,= 0.801$. The reason is that the main contribution to the emission at a given optically thin frequency comes from smaller values of $\gamma$ as the distribution of electron energies steepens (i.e., $p$ increases). With the use of $<z(p)>$ instead of the single electron value, one finds $\nu (p = 3) = 1.6\,\gamma^2 \nu_{\rm B}$ and $\nu (p = 2) = 0.95\,\gamma^2 \nu_{\rm B}$. This dependence on $p$ can be significant; for example, the magnetic field strengths deduced from inverse Compton cooling for the two different $p$-values are related by $B(p=2)/B(p=3) = 1.6$. 

Another effect that affects the relation between $\gamma$ and $\nu$ is absorption. The absorption increases faster toward lower frequencies than does the emissivity, which causes $<z>$ to decrease when absorption becomes important. The absorption coefficient is $\alpha_{\nu} \propto  \int z^{(p-2)/2} F(z) {\rm d} z$. Hence, the average value of $z(p)$ for absorption is the same as that calculated for emission with $p+1$. Since the source function is $S_{\nu} = j_{\nu}/\alpha_{\nu}$, the average value of $z$ in the part of the spectrum affected by absorption is, roughly, that for optically thin emission but calculated with $p-1$. Hence, for $p=3$, one expects that $<z(p=2)> $ should be appropriate in the absorbed part of the spectrum. As an approximation, $\nu_{\rm abs} = \gamma_{\rm abs}^2 \nu_{\rm B}$ will be used below, where $\gamma_{\rm abs}$ is the average Lorentz factor of those electrons contributing to the spectral flux at the self-absorption frequency.

\subsection{Source Properties from Light Curves as Compared to Spectra}\label{sect2d}
For a given frequency, the light curve is given by $F_{\nu} \propto R^2 [1-\exp{(-\tau)}]/B^{1/2}$, where $\tau \propto y B^{(p+6)/2} R$ is the optical depth. Let $B \propto R^{-\eta}$, which results in
\begin{equation}
	F_{\nu} \propto \left(\frac{\tau}{y}\right)^{(4+\eta)/(2-\eta(p+6))} [1 - \exp(-\tau)].
	\label{eq11}
\end{equation}
Assuming $y$ to be constant, the value of $\tau$ when the light curve peaks is obtained from
\begin{equation}
	 \exp(\tau) = 1+ \tau\frac{\{\eta (p+6)-2\}}{4+\eta}.
	\label{eq12}
\end{equation}
For $\eta = 1$ (i.e., $BR =$ constant), this yields
\begin{equation}
	\exp(\tau) = 1+ \tau \frac{(p+4)}{5}.
	\label{eq13}
\end{equation}

It can be seen that Equation (\ref{eq13}) is identical to that determining the optical depth where the spectral flux peaks \citep{bjo21}. Hence, the peak in a light curve at a given frequency $\nu$ occurs simultaneously with $\nu$ being the spectral peak. It may be noted from Equations (\ref{eq5}) and (\ref{eq6}) that this situation corresponds to $F_{\nu_{\rm abs}} = $ constant. When  $F_{\nu_{\rm abs}}$ is not constant, either due to variations in $BR$ or a time dependence in $y$, care should be taken when using light curves to obtain values of $B$ and $R$ from Equations (\ref{eq5}) and (\ref{eq6}); for example, when the value of $y$ decreases with time, one may deduce from Equation (\ref{eq11}) that the peak of the light curve occurs at an optical depth larger than that of the spectral flux. This implies that for a given frequency, its light curve peaks before it corresponds to the spectral peak.

\section{Observations of SN\,2020oi}\label{sect3}
\cite{hor20} have made detailed radio observations of SN\,2020oi. The observed spectra were fitted assuming a homogeneous source and the magnetic field and radius of the source were deduced from expressions corresponding to Equations (\ref{eq5}) and (\ref{eq6}) assuming $y=$constant. They concluded that the observed spectral flux was likely affected by inverse Compton cooling, since (1) the optically thin spectral index varied with time and (2) the time variation of the normalization of the spectra was nonstandard. The cooling frequency was then estimated from the time after which the spectra were judged to no longer be affected by cooling.

However, when cooling is important, its effects need to be included in the fitting process of the synchrotron spectra. This is most easily appreciated by considering the value of $y$. Let $y = y_{\rm o} x$, where $x = 1/(1+ t/t_{\rm cool})$ (see Equation (\ref{eq1})), so that $y_{\rm o}$ is the value of $y$ in the absence of cooling. Here, the value of $ t/t_{\rm cool}$ is calculated for $\nu = \nu_{\rm abs}$. Hence, as can be seen from Equation (\ref{eq5}), using a constant value for $y$ when cooling increases with time results in a too rapid decrease of the value of $B$, and vice versa, when the cooling subsides. Although the value of $y_{\rm o}$ is independent of cooling, its deduced value is affected by an erroneous estimate of the cooling frequency; for example, when the Compton cooling is overestimated (i.e., $\nu_{\rm cool}$ too small), the value of $B$ will be underestimated. This, in turn, increases the value of $y_{\rm o}$, i.e., one deduces a too large value for $U_{\rm rel}/U_{\rm B}$.

Ideally, the optically thin flux should have been fitted using a curved spectrum. The value of $t/t_{\rm cool}$ can then be deduced from the local spectral index (Equation (\ref{eq2})). Even so, as discussed in Section\,\ref{sect2a}, its value is likely to be quite uncertain. Furthermore, in the first few days (5 - 7), the optically thin spectral index is hard to determine due to the small observed spectral range above $\nu_{\rm abs}$. However, during days 11 and 13, the optically thin range is large enough to obtain a spectral index together with values of $\nu_{\rm abs}$ and $F_{\nu_{\rm abs}}$. Since the variation in the local spectral index is expected to be quite small, the observed value should correspond, roughly, to the middle of the observed frequency range.
 
The value of $y_{\rm o}$ is obtained directly from the observed cooling. Since $t/t_{\rm cool} = (1-x)/x$, one finds from Equations (\ref{eq9}) and (\ref{eq10}) that
\begin{equation}
	y_{\rm o}^{4/19} L_{42} = 6.7\,\frac{(1-x)}{x^{23/19}}\frac{F_{\nu_{\rm abs, 27}}^{17/19}}{t_{10}\nu_{\rm abs, 10}^2}.
	\label{eq14}
\end{equation}
This corresponds to Equation (\ref{eq8}) when cooling due to inverse Compton scattering is observed rather than the resulting X-ray emission. As long as the observables are well determined, either situation should give a good estimate of the value of $y_{\rm o}$. However, as discussed in Section\,\ref{sect2a}, in comparison to the other observables, the cooling frequency (i.e., $x$) is inherently harder to determine. Furthermore, as can be seen from Equation (\ref{eq14}), the deduced value of $y_{\rm o}$ is quite sensitive to the actual value of $x$.

A different approach would be useful in which the value of $x$ could be constrained by additional limitations on the physical variables. One such constraint would be the expectation of a roughly constant value of $y_{\rm o}$ with time. Another example is the assumed value of the peak spectral flux in the absence of cooling ($F_{\nu_{\rm abs,o}}$). Since the values of $B$ and $R$ should be unaffected by the cooling, Equations (\ref{eq5}) and (\ref{eq6}) then show that $F_{\nu_{\rm abs}} = x^{5/7} F_{\nu_{\rm abs,o}}$. Hence, if $F_{\nu_{\rm abs,o}}$ could be estimated so would the value for $x$.

An alternative scheme to constrain the source parameters is, therefore, to assume a value for $y_{\rm o}$ or $F_{\nu_{\rm abs,o}}$ and then calculate $x$ to see what range in $y_{\rm o}$- or $F_{\nu_{\rm abs,o}}$-values is allowed by observations. In principle, either one could be chosen, since once one of them is known the other can be calculated. It proves convenient to use $F_{\nu_{\rm abs,o}}$ for two reasons: (1) The value of $x$ is then given directly by $(F_{\nu_{\rm abs}} / F_{\nu_{\rm abs,o}})^{7/5}$, while calculating its value from $y_{\rm o}$ involves several observables; in particular, the initial values of  $L_{\rm bol}$ are not-so-well determined for SN\,2020oi (see Equation (\ref{eq14})). (2) Several other well-observed radio supernovae (e.g., SN\,1993J, SN 2005L, and SN\,2011dh) have roughly constant $F_{\nu_{\rm abs}}$ over rather extended periods of time, during which they are unaffected by cooling. The latter property suggests using $F_{\nu_{\rm abs,o}} = $ constant and then varying its value. 

\subsection{A Direct Way to Determine the Cooling Frequency}\label{sect3a}
The observations show that the initial decrease of $F_{\nu_{\rm abs}}$ is quite small, suggesting that the earliest measured value is close to $F_{\nu_{\rm abs, o}}$. A lower limit to the cooling is, therefore, obtained by setting $f_{\nu_{\rm abs, o}}= 5.24\,$mJy, which is the observed value at day 5. In order to estimate the effects of increased cooling on the deduced source parameters, another case with $f_{\nu_{\rm abs, o}} = 6.0\,$mJy is also calculated. The results are presented in Table\,1. Furthermore, the light-curve peak for $\nu = 5\,$GHz is not included. As discussed in Section\,\ref{sect2d}, the reason is that the spectral flux as well as the time of the peak are expected to differ from those when $\nu = 5\,$GHz corresponds to the spectral peak.

\begin{table}
\centerline{Parameters Derived for SN\,2020oi Including Cooling}
\begin{center}
\begin{tabular}{cccccccccc}
\hline\hline\\
$t$ & $F_{\nu_{\rm abs}}$ & $\nu_{\rm abs}$ & $x$ & $B\,y_{\rm o}^{4/19}$ & $R\,y_{\rm o}^{1/19}$ & $R\,y_{\rm o}^{1/19}/t$ & $B\,y_{\rm o}^{4/19} t$ & $L_{\rm bol}\,y_{\rm o}^{4/19}$ & $\Delta \alpha _{\rm 30}$\\
(days) & ($10^{27}$) & ($10^{10}$) & & & ($10^{15}$) & ($10^4$\,km/s) & (10 days) & ($10^{42}$) &\\
\toprule\\
$f_{\nu_{\rm abs, o}} = 5.24$\\
(mJy)\\[2ex]
\midrule
 5 &1.2 & 3.1&1 & 3.1& 1.9& 4.5 & 1.6 & 0 & 0\\
 6 & 1.1 & 2.4 & 0.81 & 2.5 & 2.4 & 4.6 & 1.5 & 0.51& 0.10\\
 7 & 1.0  & 2.0 & 0.78 & 2.2 & 2.8 & 4.6 & 1.5 & 0.69& 0.13\\
 11 & 0.96 &1.3 & 0.70 & 1.4 & 4.3 & 4.6 & 1.5 & 1.7 & 0.20\\
 13 & 0.90 & 0.99 & 0.65 & 1.1 & 5.4 & 4.8 & 1.4 & 2.9&0.24\\
\hline\\
$f_{\nu_{\rm abs, o}} = 6.0$\\
(mJy)\\[2ex]
\midrule
 5 &1.2 & 3.1&0.83 & 3.2& 1.9& 4.5 & 1.6 & 0.36 & 0.085\\
 6 & 1.1 & 2.4 & 0.67 & 2.6 & 2.4 & 4.7 & 1.6 & 1.1& 0.13\\
 7 & 1.0  & 2.0 & 0.65 & 2.3 & 2.8 & 4.6 & 1.6 & 1.4& 0.20\\
 11 & 0.96 &1.3 & 0.58 & 1.4 & 4.4 & 4.6 & 1.6 & 3.0 & 0.26\\
 13 & 0.90 & 0.99 & 0.54 & 1.2 & 5.4 & 4.8 & 1.5 & 4.8&0.30\\
\bottomrule
\end{tabular}
\caption{The three first columns are taken from \cite{hor20}, where the absolute spectral flux $F_{\nu}$ has been calculated   
 from the observed spectral flux $f_{\nu}$ using a distance of 14\,Mpc to SN\,2020oi.}
\end{center}
\end{table}

The values obtained for $y_{\rm o}^{4/19} L_{42}$ are shown in Table\,1. Also included is the steepening of the local spectral index due to cooling calculated for an optically thin frequency $\nu = 30\,$GHz, i.e., $\Delta \alpha_{30} = \Delta \hat{p}_{30}/2$. No error bars are included in Table\,1. The reason is that the errors given in \cite{hor20} are those resulting from fitting the observations assuming no cooling. It is not clear how they translate into the errors resulting from using instead curved optically thin synchrotron spectra appropriate for a cooling scenario. Before considering the implications of cooling, one should note that the expression to be used for $U_{\rm ph} (\propto L_{\rm bol}/R^2)$ is its time-averaged value. This depends both on the time variation of $L_{\rm bol}$ and the evolution of the shock velocity (i.e., $R$). As shown in \cite{b/f04}, when a detailed calculation was done for SN\,2002ap, the cooling (i.e., $t/t_{\rm cool}$) followed rather closely the instantaneous value of $L_{\rm bol}$, at least around its maximum. Hence, in the following, the instantaneous values of $L_{\rm bol}$ and $R$ will be used.

Before comparing the predicted steepening of the optically thin spectra ($\Delta \alpha_{30}$) to the measured values, the effects of the not-so-well-determined initial values of $L_{\rm bol}$ need to be estimated. It is useful to first consider the constraints that can be obtained from its range and then the implications of its more well-determined value around the peak. With a constant value for $y_{\rm o}$, the predicted variation in $L_{\rm bol}$ can be obtained directly from Table\,1. It can be seen that from day 6 to day 13, $L_{\rm bol}$ should have increased  by 1.9\,mag ($f_{\nu_{\rm abs, o}} = 5.24\,$mJy) or 1.6\,mag ($f_{\nu_{\rm abs, o}} = 6.0\,$mJy). The uncertainty in the early rise of the optical light curves makes both of these values consistent with observations.  However, if the indicated early rapid rise of $L_{\rm bol}$ is correct, this would favor $f_{\nu_{\rm abs, o}} = 5.24\,$mJy over $f_{\nu_{\rm abs, o}} = 6.0\,$mJy, since the latter implies strong cooling already on day 5 .

It is worth noting that a well-measured range of $L_{\rm bol}$-values would have determined the value of not only $f_{\nu_{\rm abs, o}}$ but also $y_{\rm o}$. As already mentioned, Equation (\ref{eq14}) relates the values of $F_{\nu_{\rm abs, o}}$ and $y_{\rm o}$ for a given time. With measurements at two (or more) times, their actual values can be deduced.
This shows that the observations of SN\,2020oi are such that the assumption of a constant $y_{\rm o}$-value leads to $F_{\nu_{\rm abs, o}} =$ constant and vice versa. However, there are also combinations of nonconstant values for $F_{\nu_{\rm abs, o}}$ and/or $y_{\rm o}$, which are consistent with observations. Whether or not the unique combination of both $F_{\nu_{\rm abs, o}}$ and  $y_{\rm o}$ being constants actually corresponds to reality is determined by a comparison to the observed spectral curvature. It should also be emphasized that the possibility of both $F_{\nu_{\rm abs, o}}$  and $y_{\rm o}$ being constants is not something that can be generally assumed but requires special source properties.

The cooling induced spectral steepening on days 11 and 13 are 0.20 and 0.24 ($f_{\nu_{\rm abs, o}} = 5.24\,$mJy) and 0.26 and 0.30 ($f_{\nu_{\rm abs, o}} = 6.0\,$mJy), respectively. With $p=3$, the corresponding values in \cite{hor20} are 0.17 and 0.20 (their Figure 9). Since the formal errors are, roughly, 0.05, a straightforward comparison would then select $f_{\nu_{\rm abs, o}} = 5.24\,$mJy. As already mentioned, the actual errors are hard to estimate, and hence, a strong conclusion cannot be reached. However, together with the above discussion of the initial variations of $L_{\rm bol}$, this shows that an inverse Compton cooling scenario is fully consistent with the observations of SN\,2020oi and that $f_{\nu_{\rm abs, o}} = 5.24\,$mJy is to be preferred over $f_{\nu_{\rm abs, o}} = 6.0\,$mJy. 

\section{Discussion}\label{sect4}
The value of $y_{\rm o}$ can be derived from the values of $y_{\rm o}^{4/19}L_{\rm bol, 42}$ given in Table\,1. The various optical light curves peaked around days 11 and 13. In this time range, \cite{hor20} used $L_{\rm bol, 42} \approx 2.3$ to calculate the X-ray emission. With this value of $L_{\rm bol}$, one finds that $f_{\nu_{\rm abs, o}} = 5.24\,$mJy implies $y_{\rm o} \approx 1$.  Due to the sensitivity of $y_{\rm o}$ to the actual value of $y_{\rm o}^{4/19}L_{\rm bol, 42}$,  $f_{\nu_{\rm abs, o}} = 6.0\,$mJy  suggests a $y_{\rm o}$-value at least an order of magnitude larger than for $f_{\nu_{\rm abs, o}} = 5.24\,$mJy.

The effects of cooling on the time evolution of $B$ and $R$ can be directly seen in Table\,1. With $y_{\rm o}$ independent of time, one notices that the observations are consistent with $B \propto t^{-1}$. This conclusion differs from that in \cite{hor20}, since they did not include cooling in the synchrotron fitting procedure. Hence, there is no need to invoke a varying mass-loss rate from the progenitor star prior to the supernova explosion.

The effects of an inhomogeneous source structure can also be estimated. When self-absorption is important, the source filling factor ($\phi_{\rm fil}$) needs to be split into the source covering factor ($\phi_{\rm cov}$) and the reduced line-of-sight extension of the source ($\phi_{\rm los}$), so that $\phi_{\rm fil} = \phi_{\rm cov}\times \phi_{\rm los}$. The values of $y_{\rm o}^{4/19}L_{\rm bol, 42}$ in Table\,1 assume $\phi_{\rm cov}=1$. When this is not the case, $F_{\rm abs}$ should be replaced by $F_{\rm abs}/\phi_{\rm cov}$, and hence, $y_{\rm o} \propto \phi_{\rm cov}^{-17/4}$ (see Equation (\ref{eq14})). Inhomogeneities along the line of sight do not affect the value of $y_{\rm o}$ but rather its relation to the other source parameters, since $R_{||}$ should then be replaced by $R_{||} \times \phi_{\rm los}$ (see Equation (\ref{eq7})). Thus, the value of $\gamma_{\rm min} U_{\rm rel}/U_{\rm B}$ is much more sensitive to $\phi_{\rm cov}$ than $\phi_{\rm los}$.

The lack of an observed low energy cutoff in the electron distribution constrains $\gamma_{\rm min}$ to be smaller than $\gamma_{\rm abs}$. Since $\gamma_{\rm abs} \approx 60$ (see Equation (\ref{eq5})) for supernovae and $R_{||}/R = 1/2$,  a value of $U_{\rm rel}/U_{\rm B} > 1$ can be claimed only for $y_{\rm o}\,\gsim\,30\,\phi_{\rm los}$. Hence, it is seen that in order for this to be the case,  either the source needs to be inhomogeneous or the cooling should be at least as strong as that corresponding to $f_{\nu_{\rm abs, o}} = 6.0\,$mJy. However, as noted above, the observations fit better with  $f_{\nu_{\rm abs, o}} = 5.24\,$mJy than $f_{\nu_{\rm abs, o}} = 6.0\,$mJy. Therefore, one may conclude that the simplest interpretation of the observations is that the radio source in SN\,2020oi is homogeneous and that equipartition between relativistic electrons and magnetic field applies. 

It is usually assumed that the non-thermal electron distribution is due to first-order Fermi acceleration at a shock front. However, before this process starts to be effective for the electrons, they need to be pre-accelerated in order for them to experience the whole pressure increase across the shock front. The width of the shock front is roughly the mean-free path of the thermal ions/protons, which in the Bohm approximation corresponds to their Larmor radius. Hence, Fermi acceleration of the electrons is expected to begin when their Lorentz factor exceeds $(m_{\rm p}/m_{\rm e})(v/c)$, where $m_{\rm p}/m_{\rm e}$ is the mass-ratio between ion/protons and electrons. This scenario is confirmed in detailed particle-in-cell calculations, and for example, \cite{par15} find that electrons are injected into the Fermi-acceleration process when their Larmor radii are a few times that of the ions/protons. However, in supernovae, $v/c \sim 0.1$, so that the corresponding Lorentz factor is a few times $10^2$. This is substantially larger than the value of $\gamma_{\rm abs}\,(\approx 60)$, and hence, it is likely that the Fermi accelerated part of the electron distribution is never observed.

Although first-order Fermi acceleration is well understood, the pre-acceleration phase is not. The observed optically thin spectra in supernovae indicate $p\approx 3$ \citep[e.g.,][]{c/f06}. This is often taken to indicate that the standard scenario for first-order Fermi acceleration needs to be modified, since the canonical value is $p=2$. However, since the radio emission in supernovae is likely to come from electrons in the pre-acceleration phase, this gives instead an opportunity to directly study the physical mechanisms leading up to Fermi acceleration. This involves not only the distribution of electron energies but, more importantly, the injection efficiency and the value of $\gamma_{\rm min}$. One may note that the simulations done by \cite{par15} gave $p\approx 2$ also in the pre-acceleration phase.

When cooling is included, the variation of the magnetic field strength is consistent with $B\propto t^{-1}$, which conforms to the common assumption that the energy density of the magnetic field should scale with the thermal energy density behind the shock. However, with a constant peak spectral flux, this implies $R\propto t$, i.e., a constant velocity. A roughly constant velocity of the forward shock implies a very steep density gradient of the ejecta in order to provide the needed momentum input without changing the ejecta velocity at the reverse shock too much. On the other hand, the deduced value of the magnetic field gives a lower limit to the energy density in the synchrotron emitting region, and hence, the density of the circumstellar medium. Since the modeling of supernova explosions gives a rather limited range for the total kinetic energy of the ejecta, the large value of $B$ suggests a large amount of kinetic energy at high ejecta velocities, which, in turn, implies a shallow density gradient of the ejecta. These two constraints cannot always be simultaneously met. An alternative is to assume that the value of the break in the ejecta velocities ($v_{\rm o}$) is substantially larger than expected.

The implications of the observations of SN\,2020oi can be quantified by assuming the density of the ejecta to scale with radius ($r$) as $\rho_{\rm ej} \propto r^{-\rm n}$. The velocity of the forward shock then varies with time as $v \propto t^{-\rm{1/(n-2)}}$ \citep{che82a}. Although the velocities given in Table\,1 (i.e., $R/t$) are consistent with being constant, errors are such that a decrease of 4\%-5 \% cannot be excluded between days 5 and 13; this implies $n\,\gsim\,25$. As shown in the Appendix, this leads to $v_{\rm o}\,\gsim\,3.8\times 10^4$\,km/s (and a corresponding ejecta mass $M_{\rm ej}\,\lsim\,6.4\times10^{-2} \Msun$). This is a rather extreme value for a standard supernova explosion. One may also note that a covering factor less than unity would lead to an even more extreme value. 

This is similar to the situation in SN\,1993J \citep{f/b98}, where the kinetic energy at high ejecta velocities, needed to account for the roughly constant velocity during the first few hundred days, was more than an order of magnitude larger than even the most favorable model could provide. In SN\,1993J, the velocity of the outer rim of the emission region was not deduced by model fitting but directly measured from VLBI-observations \citep{bar94,mar95a,mar95b}. It was suggested in \cite{bjo15} that the magnetic field in SN\,1993J was amplified by the Rayleigh-Taylor instability at the contact discontinuity, and hence, driven by the reverse shock. Initially, then, the emission region grew outward from the contact discontinuity until it saturated when approaching the forward shock \citep{che92}. As a result, the observed, roughly constant velocity of the outer rim was a combination of a growing emission region and a decelerating ($n\approx 7$) outer shock. 

Such an origin for a roughly constant velocity of the source outer radius leads to an increasing value of $F_{\nu_{\rm abs}}$. Hence, a similar scenario cannot be invoked for SN\,20120oi, since, here, the peak spectral flux decreased with time. It can be seen from Equations (\ref{eq5}) and (\ref{eq6}) that the product $BR$ is independent of $\nu_{\rm abs}$, while their individual values are quite sensitive to $\nu_{\rm abs}$; for example, $R \propto \nu_{\rm abs}^{-1}$. Since the large values implied for $n$ are a direct result of the time evolution of $\nu_{\rm abs}$, it is worth considering the possibility that its decrease with time is over-estimated. Since normally $\nu_{\rm abs}$ is deduced from observations by fitting spectra appropriate for a homogeneous source structure, there are at least two situations where this can systematically affect the value of the peak frequency. In an inhomogeneous source, the spectral peak is broadened by the inhomogeneities \citep{b/k17}. The fitting procedure is then not straightforward and can lead to rather large uncertainties \citep[e.g.,][]{sod12,ale15}; in particular, inhomogeneities varying with time may cause systematic effects. 

As argued above, apart from cooling, the inhomogeneities in SN\,2020oi are likely to be small. However, cooling results in spectral curvature at optically thin frequencies. \cite{hor20} accounted for the varying cooling by changing the optically thin spectral index. For a given optically thick part, the value of $\nu_{\rm abs}$ decreases for steeper optically thin spectra (i.e., larger $p$). Hence, during times when cooling increases, fitting observations with homogeneous spectra will systematically overestimate the rate of decrease in $\nu_{\rm abs}$, and vice versa, when the cooling decreases. Since the spectra in \cite{hor20}, for which the spectral peak could be observed, occurred during the period when cooling was increasing, this will artificially increase the decline rate of $\nu_{\rm abs}$. 

Although the magnitude of this effect is hard to estimate, it illustrates that for sources in which cooling and/or inhomogeneities are indicated, the actual fitting procedure of the observations can seriously affect some of its deduced parameters. The parameter that is most sensitive is $n$, since its value determines, in large part, the properties of the supernova ejecta. As an example, for a progenitor star with radiative envelope, which is thought appropriate for SN\,2020oi, \cite{m/m99} derived $n=10.18$. In the case of SN\,2020oi, this leads to  $v_{\rm o} =1.4 \times 10^4$\,km/s ($M_{\rm ej} = 3.6\times10^{-1} \Msun$, see the Appendix). This differs substantially from the value derived above and is in the range expected at least for SNe Ib \citep [e.g.,][]{woo19}. Therefore, it would be interesting to estimate the probability that $\nu_{\rm abs} \propto t^{-0.88}$ (corresponding to $n=10.18$) is compatible with the observations of SN\,2020oi.

\section{Conclusions}\label{sect6}
The main point of the present paper is that when cooling is important, it needs to be included in the fitting process directly rather than estimating its effects afterward. Neglecting to do so can drastically affect the deduced source parameters. For SN\,2020oi, it is shown that contrary to previous claims:

1) The observations are consistent with equipartition between relativistic electrons and magnetic field.

2) There is no need to invoke a varying mass-loss rate of the progenitor star prior to the supernova explosion.

In addition, attention is drawn to a few implications for supernovae in general from the observations of SN\,2020oi:

3) In the standard first-order Fermi-acceleration scenario, radio emission in supernovae is likely due to electrons in their pre-acceleration phase, i.e., before they enter the Fermi acceleration.

4) It is important to determine the time evolution of the synchrotron self-absorption frequency, since it is directly related to the properties of the ejecta. The deduced, slow deceleration of the forward shock in SN\,2020oi indicates that a large fraction of the total kinetic energy resides at large ejecta velocities. 

5) In the absence of cooling, SN\,2020oi would have had a constant peak spectral flux over an extended period of time. This is in line with several other well-observed supernovae.

{\bf Acknowledgements:} Thanks are due to Anders Jerkstrand for helpful comments regarding the ejecta structure of stripped-envelope supernovae.

\newpage

\appendix

\begin{center}
{\bf Appendix}
\end{center}

\section{The ejecta structure in supernovae}
With a spherically symmetric supernova explosion, the density of the ejecta at a time $t$ can be written
\begin{equation}
	\rho_{\rm ej} = \frac{(n-3)M_{\rm ej}}{4\pi(v_{\rm o}t)^3} \left(\frac{v}{v_{\rm o}}\right)^{-n},
	\label{eqA1}
\end{equation}
where $M_{\rm ej}$ is the total ejecta mass for ejecta velocities larger than $v_{\rm o}$, where $v_{\rm o}$ corresponds to the break in the velocity distribution. Furthermore, the radius ($r$) is related to the ejecta velocity ($v$) through $r = vt$ and it is assumed that the maximum ejecta velocity is much larger than $v_{\rm o}$. In a spherical wind from the progenitor star with velocity $v_{\rm w}$ and a constant mass-loss rate $\dot M$, the density in the wind is $\rho_{\rm w} =\dot M/(4\pi v^2t^2v_{\rm w})$.

The interaction between ejecta and wind has a self-similar structure in which $\rho_{\rm ej} = x(n) \rho_{\rm w}$ \citep{che82a}, where, in the thin shell approximation, $x(n) = (n-3)(n-4)/2$ \citep{che82b}. This can be rewritten as
\begin{equation}
	t = \frac{2}{n-4}\frac{M_{\rm ej}}{v}\frac{v_{\rm w}}{\Mdot}\left(\frac{v}{v_{\rm o}}\right)^{(3-n)},
	\label{eqA2}
\end{equation}
where $v$ is now the ejecta velocity at the reverse shock. The total kinetic energy of the ejecta is
\begin{equation}
	E_{\rm k} =\frac{(n-3)}{(n-5)}\frac{M_{\rm ej}v_{\rm o}^2}{2}.
	\label{eqA3}
\end{equation}

The ejecta properties are described by $v_{\rm o}$ and $M_{\rm ej}$. From Equations (\ref{eqA2}) and (\ref{eqA3}) one finds
\begin{equation}
	\left(\frac{v}{v_{\rm o}}\right)^{(n-5)} = 7.4\times 10^5 \frac{(n-5)}{(n-3)(n-4)} \frac{E_{\rm k,51}}{v_{\rm 9}^3 t_{\rm 10}}
	\frac{v_{\rm w,8}}{\Mdot_{-5}},
	\label{eqA4}
\end{equation}
where $E_{\rm k,51} \equiv E_{\rm k}/10^{51}$, $v_{\rm 9} \equiv v/10^9$, $v_{\rm w,8} \equiv v_{\rm w}/10^8$, $t_{\rm 10} \equiv t/10$\,days, and $\Mdot_{-5} \equiv \dot M /(10^{-5}\,M_{\odot}/{\rm yr)}$. The thermal energy density behind the forward shock is
\begin{equation}
	U_{\rm th} = 7.6\times 10^{-1}\frac{\Mdot_{-5}}{v_{\rm w,8}\,t_{10}^2}.
	\label{eqA5}
\end{equation}
As argued in Section\,\ref{sect4}, $y_{\rm o} \approx 1$ and together with the canonical assumption that $U_{\rm B}/U_{\rm th} \approx 0.1$, the values in Table\,1 indicate $\Mdot_{-5}/v_{\rm w,8} \approx 1$. Furthermore, the ejecta velocity at the reverse shock is somewhat higher than the velocities given in Table\,1, which correspond to that of the forward shock \citep{che82a}, so that $v_{\rm 9}\approx 5$. Hence, with $E_{\rm k,51} =1$, $v_{\rm 9} = 5$, $t_{\rm 10} = 1$, and $\Mdot_{-5}/v_{\rm w,8} = 1$, one finds from Equation (\ref{eqA4}) for $n = 25$ that $v/v_{\rm o} = 1.3$, or $v_{\rm o,9} = 3.8$, and  from Equation (\ref{eqA3}) $M_{\rm ej} = 6.3\times10^{-2} \Msun$. If, instead, $n = 10.18$ is used, one derives $v/v_{\rm o} = 3.5$, or $v_{\rm o,9} = 1.4$, and  from Equation (\ref{eqA3}) $M_{\rm ej} = 3.6\times10^{-1} \Msun$.

\end{document}